\documentclass[journal]{IEEEtran}
\usepackage{graphicx}
\usepackage{cite}
\usepackage{picinpar}
\usepackage{amsmath}
\usepackage{url}
\usepackage{flushend}
\usepackage[latin1]{inputenc}
\usepackage{colortbl}
\usepackage{soul}
\usepackage{multirow}
\usepackage{pifont}
\usepackage{color}
\usepackage{alltt}
\usepackage{hyperref}
\hypersetup{
	colorlinks = true, linkcolor = magenta, anchorcolor = magenta,
	citecolor = magenta, filecolor = orange, urlcolor = orange}
\usepackage{enumerate}
\usepackage{epstopdf}
\usepackage{pbox}
\usepackage{enumitem}
\usepackage{fancyhdr}
\usepackage[switch,pagewise]{lineno}
\modulolinenumbers[2]

\usepackage{bm}
\usepackage{amssymb}
\usepackage{stfloats}
\usepackage{algorithmic}
\usepackage[ruled,vlined]{algorithm2e}
\usepackage[table]{xcolor}

\newcommand{\rv}[1]{\textcolor{black}{#1}}

\usepackage{wasysym}
\usepackage{diagbox}
\begin{document}

\title{Secure Blockchain Platform for Industrial IoT with Trusted Computing Hardware}

\author{Qing~Yang,
        Hao~Wang,
        Xiaoxiao~Wu,
        Taotao~Wang,
        Shengli~Zhang,
        Naijin~Liu
\thanks{This work is in part supported by the National Natural Science Foundation of China (project 61901280) and the FIT Academic Staff Funding of Monash University. (Corresponding author: Hao Wang.)}
\thanks{Q.~Yang, X.~Wu, T.~Wang, and S.~Zhang are with the Blockchain Technology Research Center (BTRC) and the College of Electronics and Information Engineering (CEIE), Shenzhen University, Shenzhen, Guangdong Province, PRC, e-mail: {yang.qing@szu.edu.cn}, {xxwu.eesissi@szu.edu.cn}, {ttwang@szu.edu.cn}, {zsl@szu.edu.cn}.}
\thanks{H. Wang is with the Department of Data Science and Artificial Intelligence, Faculty of Information Technology, Monash University, Melbourne, VIC 3800, Australia, e-mail: hao.wang2@monash.edu.}
\thanks{Naijin~Liu is with the Qian Xuesen Space Technology Laboratory, China Academy of Space Technology, Beijing, China, e-mail: {liunaijin@qxslab.cn}.}
}

\maketitle

\begin{abstract}
As a disruptive technology that originates from cryptocurrency, blockchain provides a trusted platform to facilitate industrial IoT (IIoT) applications. However, implementing a blockchain platform in IIoT scenarios confronts various security challenges due to the rigorous deployment condition. To this end, we present a novel design of secure blockchain based on trusted computing hardware for IIoT applications. Specifically, we employ the trusted execution environment (TEE) module and a customized security chip to safeguard the blockchain against different attacking vectors. Furthermore, we implement the proposed secure IIoT blockchain on the ARM-based embedded device and build a small-scale IIoT network to evaluate its performance. Our experimental results show that the secure blockchain platform achieves a high throughput (150TPS) with low transaction confirmation delay (below 66ms), demonstrating its feasibility in practical IIoT scenarios. Finally, we outline the open challenges and future research directions.
\end{abstract}


\markboth{IEEE Internet of Things Magazine}{}

\section{Introduction}\label{s:intro}
\IEEEPARstart{R}{ecent} years have witnessed the wide application of the Internet of Things (IoT) technology in the industrial context by connecting massive smart IoT devices. The rapid development of Industry~4.0 brings both opportunities and challenges to the industrial IoT (IIoT) scenarios such as automatic manufacturing, smart logistics, industrial sensor network, and fog computing \cite{cano2018evolution}. To facilitate these IIoT applications, a trusted platform is needed to provide trusted data acquisition, identity management, and decentralized computing. Blockchain, being the disruptive technology underpinning the cryptocurrency, provides a promising solution to the above issues.

\begin{figure*}[!t]
    \centering
    \includegraphics[width=12cm]{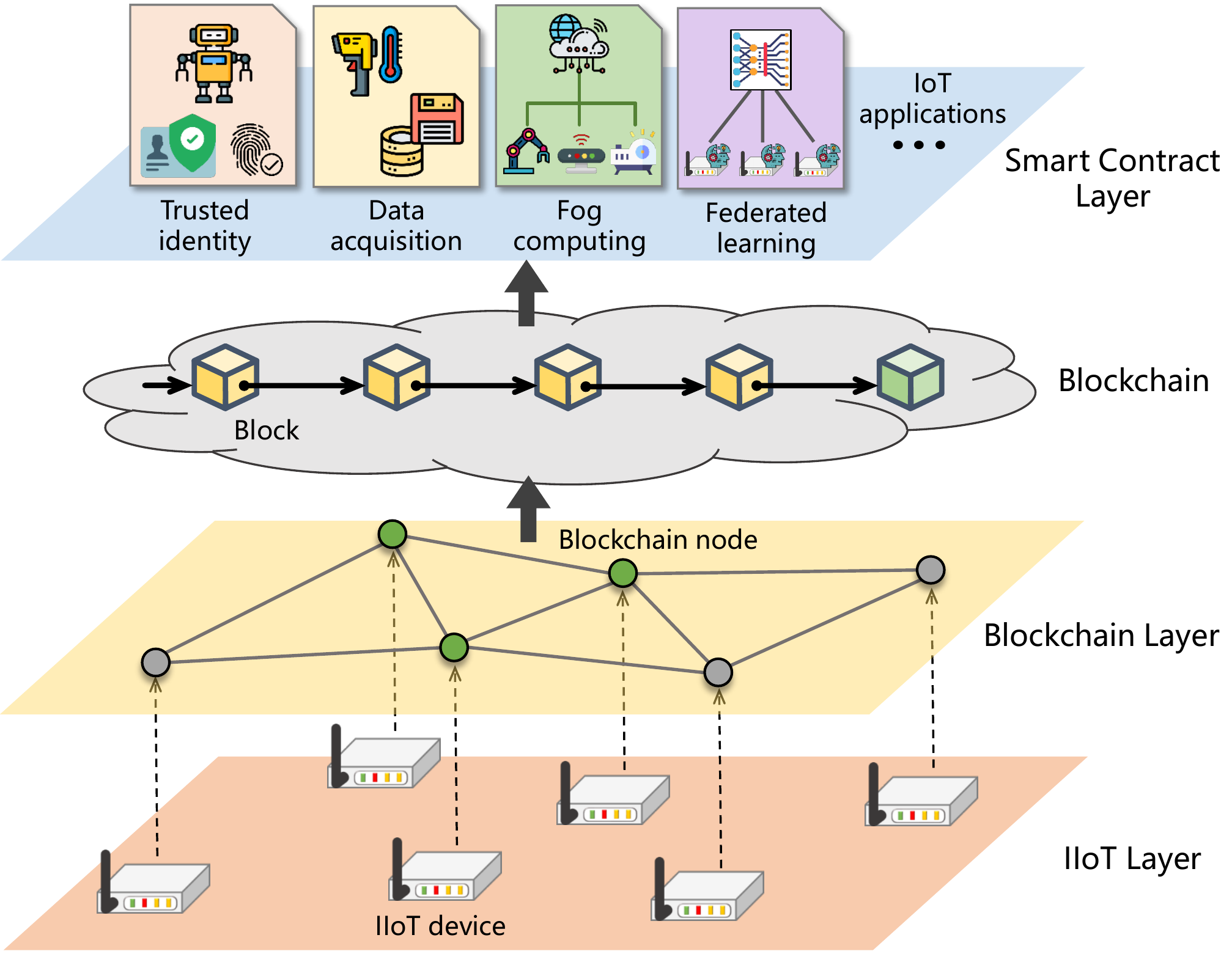}
    \caption{The architecture of the blockchain platform for industrial IoT applications.}
    \label{f1}
\end{figure*}

Blockchain is a tamper-proof decentralized ledger maintained by a group of nodes through the consensus algorithm \cite{dinh2018untangling}. Furthermore, the blockchain supports the execution of generic computer programs as smart contracts, resulting in the proliferation of various decentralized applications \cite{eth}. We run the blockchain software on the IIoT devices and connect them to form the IIoT blockchain via IoT communication technologies such as Narrowband-IoT and LoRa \cite{ferrag2018blockchain}. Fig.~\ref{f1} depicts the architecture of the IIoT blockchain platform that consists of the IIoT layer, the blockchain layer, and the smart contract layer. The IIoT blockchain can support various IIoT applications on top of the smart contract layer as a decentralized computing platform.

However, building such a blockchain platform in a practical IIoT system confronts critical security challenges. In industrial scenarios, the IIoT devices are exposed in open areas, thus vulnerable to hostile attacks such as data tampering, system hacking, and identity spoofing. Even worse, IIoT devices are usually designed for long-time operation (e.g., sensors), so once launched there is no effective method to fix the software or hardware security issues. To address this challenge, we advocate that trusted computing hardware \cite{smith2013trusted} should be leveraged to lay the security foundation for the IIoT blockchain. This work presents the design and implementation of a secure blockchain platform based on the trusted computing hardware for IIoT scenarios. The main contributions of this paper are as follows.
\begin{itemize}
    \item We present a comprehensive design of a high-performance blockchain for IIoT devices from the aspects of membership selection, consensus algorithm, and blockchain structure.
    \item We propose a novel framework of secure IIoT blockchain based on trusted computing hardware. Specifically, we utilize the trusted execution environment (TEE) and a customized security chip to safeguard the blockchain platform from various malicious attack vectors.
    \item We implement the proposed secure blockchain on practical IIoT devices to validate its feasibility and build a testing IIoT network to evaluate its performance.
\end{itemize}

The remainder of the article is organized as follows. Section~\ref{s:model} describes the design of the IIoT blockchain. Section~\ref{s:secure} presents the architecture of the secure blockchain platform based on trusted computing hardware. Section~\ref{s:implementation} implements and evaluates the proposed secure blockchain on practical IIoT devices. Section~\ref{s:future} discusses future research directions and Section~\ref{s:conclusion} concludes the paper.

\section{Blockchain for Industrial IoT}\label{s:model}

This section presents the design of the proposed secure blockchain for IIoT. \rv{We begin with a brief introduction to blockchain from the data processing perspective; then, we elaborate on the modularity of our blockchain into three modules of membership selection, consensus algorithm, and blockchain structure.}

\subsection{Blockchain in a Nutshell}\label{ss:bc}

The word ``Blockchain'' is first found in the whitepaper of Bitcoin, but its concept originates from the research area of distributed computing \cite{dinh2018untangling}. As shown in Fig.~\ref{f1}, the IIoT blockchain is a network of nodes in which each node corresponds to an IIoT device. The blockchain can be regarded as a state machine that records the states of all the nodes. For example, in Bitcoin, the state of a node is the Bitcoin balance of the node, and the Bitcoin blockchain is a global ledger that stores the balances of all the nodes. Each blockchain node keeps a replica of the global state, and all the nodes maintain the state machine in a decentralized manner thus removes the need for a central server.

A \emph{transaction} is a set of operations that can change the global state of the blockchain. For example, a Bitcoin transaction containing Bitcoin transfer from user A to user B will change the states (e.g., balances) of both users A and B. \rv{Multiple transactions are batched into a \emph{block} to increase the processing efficiency and lower the communication overhead.} For example, one Bitcoin block contains around 2,000 transactions. In an ideal scenario, all the blockchain nodes start in the same initial state, execute the operations of the transactions in the same order, and reach the same next state synchronously.

However, a practical blockchain system confronts challenges such as asynchronous networks, hardware breakdown, and even malicious nodes. To build a high-performance IIoT blockchain in practice, we focus on the three key components of the blockchain, namely, membership selection, consensus algorithm, and chain structure.

\begin{figure*}[!htb]
    \centering
    \includegraphics[width=14cm]{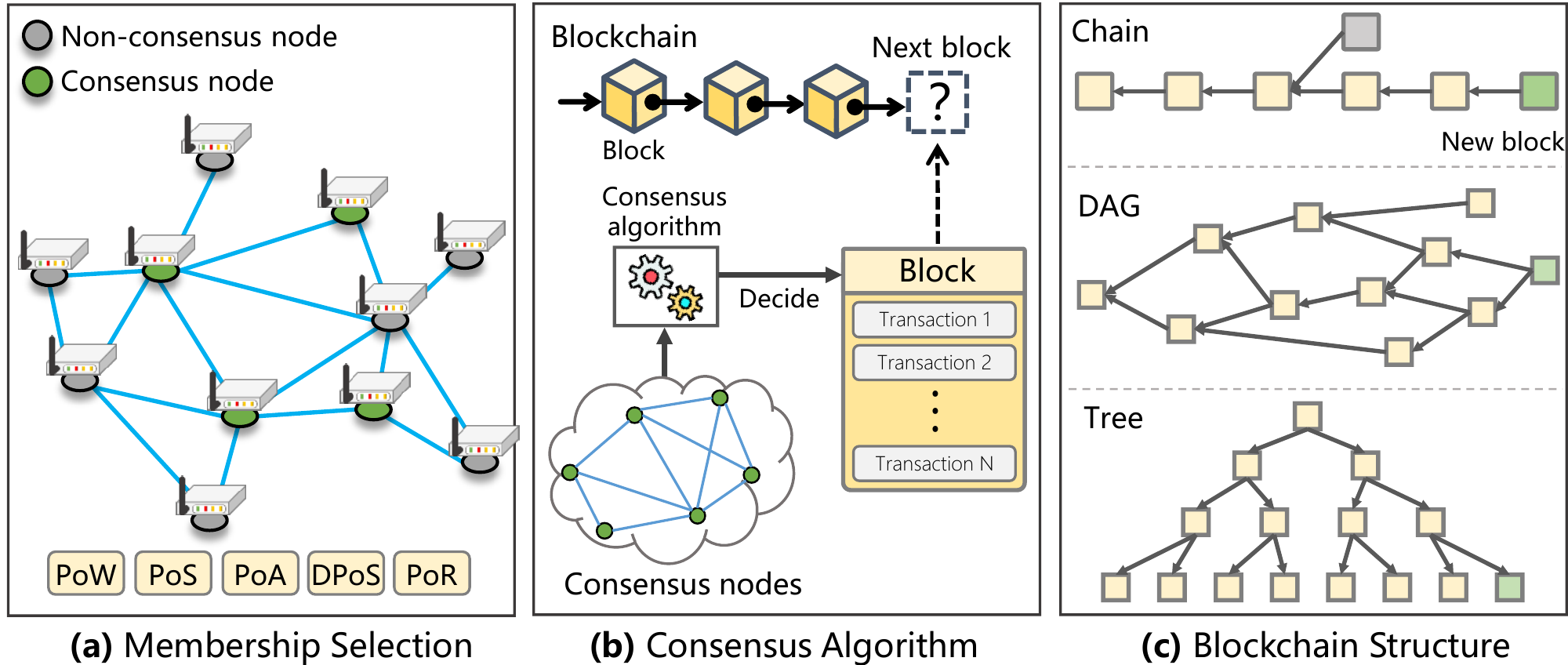}
    \caption{The key components of the IIoT blockchain: (a) membership selection; (b) consensus algorithm; (c) chain structure.}
    \label{f2}
\end{figure*}

\subsection{Membership Selection}\label{ss:membership}

\rv{We employ the consortium blockchain considering the network scale and the performance requirement of the IIoT applications.} As shown in Fig.~\ref{f2}a, we categorize the blockchain nodes into two types: the consensus node and the non-consensus node. The consensus nodes can execute the transactions and change the state of the blockchain, while the non-consensus nodes can only verify and accept the state change. The function of the membership selection is to determine a group of consensus nodes that maintains the blockchain using the consensus algorithm. With the proliferation of crypto-currency, many membership selection methods have been proposed, such as proof-of-work (PoW), proof-of-stake (PoS), and proof-of-authority (PoA). In PoW, any node can become the consensus node if it can solve a puzzle that requires the tremendous computation of hash functions. In PoS, a node can become the consensus node if it stakes something of value on the blockchain, such as the token. In PoA, a pre-defined committee of consensus nodes is selected to maintain the blockchain.

\rv{We employ the PoS as the membership selection method for the IIoT blockchain because PoS is more suitable for IIoT devices based on the following considerations. First, the PoS is feasible on IIoT devices because of its low computational complexity. By contrast, solving the puzzle in PoW consumes a large amount of energy and computational power, which is infeasible for IIoT devices.} Second, the PoS mechanism is flexible and easy to scale. Unlike the PoA that selects a fixed group of consensus nodes, PoS can adjust the size of the consensus group to include more eligible nodes, which is suitable for large-scale IIoT networks. Third, the robustness of PoS has been proved by mainstream blockchain projects such as EOS and ETH 2.0.

\subsection{Consensus Algorithm}\label{ss:consensus}
The consensus algorithm enables the consensus nodes to cooperatively determine the next state of the blockchain in the presence of malicious nodes. As shown in Fig.~\ref{f2}b, all the consensus nodes must reach an agreement on the order of transactions included in the block so that all the nodes move to the same state. However, there may exist malicious consensus nodes, also known as the Byzantine nodes, that behave arbitrarily to sabotage the consensus process. Therefore, the consensus algorithm must be able to tolerate the Byzantine nodes as well as other adverse factors in the IIoT scenario such as device breakdown and network delay.

In this work, we implement a simplified practical Byzantine fault tolerance (PBFT) consensus algorithm for the IIoT blockchain. A committee of $N$ validators is selected using the PoS mechanism described in Section~\ref{ss:membership} to run the consensus algorithm. We divide the consensus process into epochs, and one consensus epoch consists of three phases: \emph{PREPARE}, \emph{PRE-COMMIT}, and \emph{COMMIT}. At the beginning of each consensus epoch, one validator is assigned to be the leader in a round-robin manner. In the \emph{PREPARE} phase, the leader collects the transactions, packages them into a new block, and broadcasts the block to other validators who will then verify and vote for the block.  In \emph{PRE-COMMIT} and \emph{COMMIT} phases, the leader waits for $(N-f)$ votes, where $f$ is the number of Byzantine nodes, from other validators; upon receiving enough votes, the leader aggregates all the votes into a single certificate, broadcasts it to other validators, and moves to the next phase. At the end of the COMMIT phase, all the validators can reach a consensus on the proposed new block and begin the next consensus epoch. This consensus algorithm reduces the communication complexity of the original PBFT algorithm \cite{castro1999practical} from $O(N^2)$ to $O(N)$, which saves communication bandwidth and provides higher throughput.

\subsection{Blockchain Structure}
The structure of the blockchain is also a vital component that impacts the performance of the IIoT blockchain system. Fig.~\ref{f2}c illustrates three popular blockchain structures used by mainstream blockchain projects. The \emph{chain} structure organizes the blocks in a single chain, and the new block is always appended to the end of the chain. The chain structure is simple to implement and manage but has the risk of forking. The \emph{directed acyclic graph} (DAG) structure organizes the blocks as the vertices of the DAG, thus avoiding the problem of forking. The DAG structure allows multiple consensus nodes to append new blocks to the blockchain simultaneously hence achieves higher throughput; therefore, the DAG structure is adopted by some recent high-performance blockchains, such as IOTA and Conflux. The \emph{tree} is another efficient data structure to organize the blocks. The Libra blockchain stores all the transactions in the Merkle tree, which provides efficient proof-of-existence and retrieval for transactions.

\rv{In this work, we adopt the chain-structure blockchain for higher throughput, faster transaction confirmation, and better support for smart contracts. Additionally, the forking problem of the chain structure is avoided by using the simplified PBFT consensus algorithm; therefore, adopting the classical chain structure also eases the system implementation. Different modules of the secure blockchain are well decoupled and can be reused or replaced. We follow the block structure design of the Ethereum \cite{geth} except that we add an extra data field in the block header to store the aggregated PBFT certificate.}

\section{Secure Blockchain Platform Based on Trusted Computing Hardware}
\label{s:secure}
\begin{figure*}[!t]
    \centering
    \includegraphics[width=12cm]{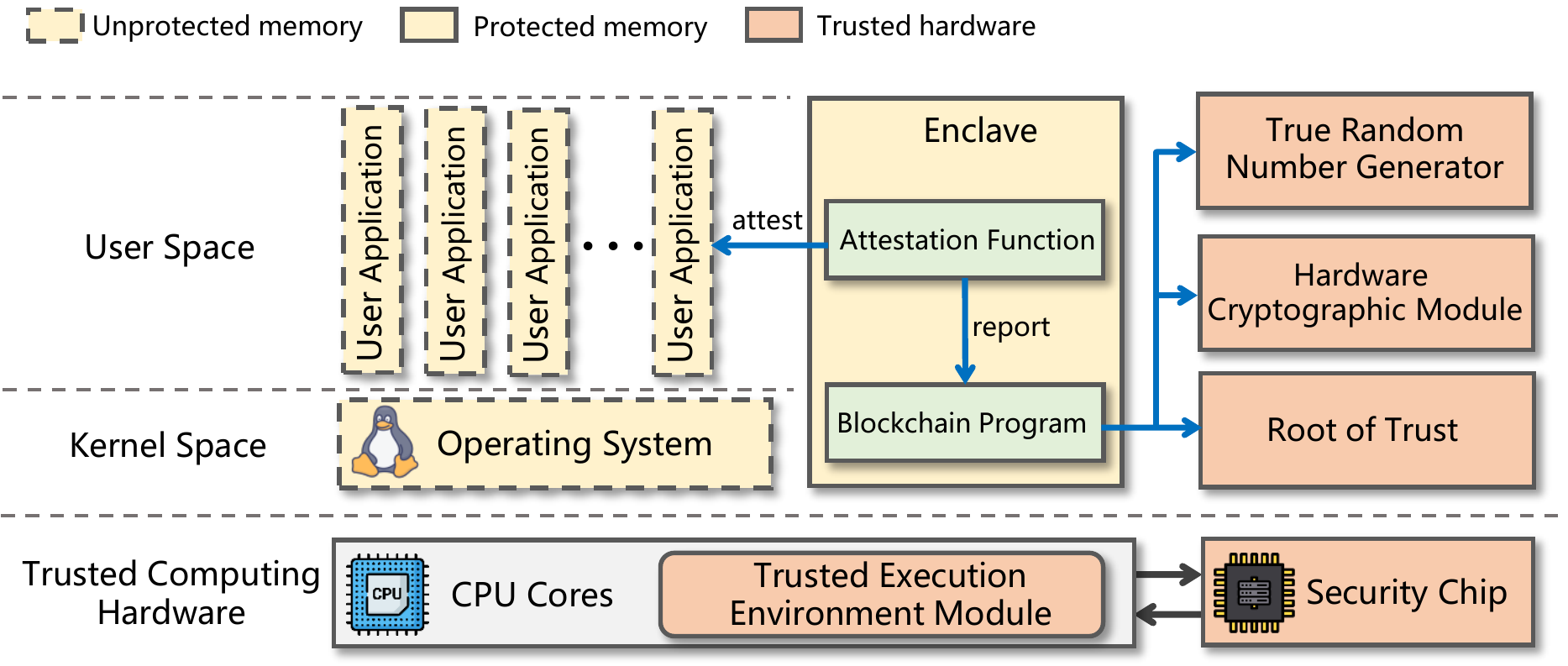}
    \caption{The secure blockchain platform based on trusted computing hardware.}
    \label{f3}
\end{figure*}

This section describes the architecture of the secure blockchain platform for IIoT. We first introduce security threats that may undermine the trust of the blockchain platform. We then present our approach to cope with the security threats using TEE and a customized security chip. Finally, we discuss the software framework of the secure blockchain platform based on trusted computing hardware.

\subsection{Threat Model}
\rv{As a blockchain node, the IIoT device is exposed to the harsh industrial environment during all its life cycle; therefore, the blockchain node is vulnerable to various security threats. In this work, we consider four types of attack vectors that an IIoT device may confront as follows.}
\begin{itemize}
    \item \textbf{Hardware attack}: \rv{the attacker can intercept, modify, or replay any electric signal outside of the chip package on the circuit board of the IoT device. These signals include the data and control messages on the bus as well as the information exchanged between chips.} 
    \item \textbf{Software attack}: the attacker can hack the operating system (OS), hijack the user-space application, and control the network traffic. We also assume that the attacker can arbitrarily read and modify any memory that is unprotected by the trusted computing hardware.
    \item \textbf{Side-channel attack}: \rv{the attacker can glean information by monitoring the cached content, measuring the timing signal of the chips, and observing the information on the data/control channel of the device.}
    \item \textbf{Denial-of-service (DoS) attack}: the attacker can take down the OS and user-space applications with the DoS attack, but cannot affect the trusted computing hardware and the applications inside the TEE enclave.
\end{itemize}

\rv{To support the functionality of the secure blockchain system, the following requirements must be met. First, the CPU of the target IoT device should support the TEE module, e.g., the TrustZone or SGX. Second, we assume that the attacker cannot pry into the chips since it will damage the hardware of the blockchain node. Third, the IoT devices have stable network connections, such as Ethernet or 5G NB-IoT.}

\subsection{Trusted Computing Hardware}
To counter the above attack vectors, we employ the trusted computing hardware that consists of the trusted execution environment (TEE) inside the CPU cores and the customized security ASIC chip, as shown in Fig.~\ref{f3}. The TEE is a secure area of the CPU core that can create an isolated piece of memory called "enclave" \cite{lee2020keystone}. To further strengthen the security guarantee, we build an ASIC chip that provides security-critical functions at the hardware level. The security chip runs in a standalone mode that cannot be intervened by any outsiders. The trusted computing hardware \rv{lays} the foundation for the trusted software framework described in the next section. We elaborate on the principle of the trusted computing hardware as follows.

\subsubsection{Isolated execution}
We use the TEE module to isolate the execution of security-critical programs on the IIoT device. Mainstream CPU vendors respectively provide their implementation of the TEE module, such as the TrustZone from ARM, SGX from Intel, and SEV from AMD. To create an enclave, the TEE module claims a piece of memory during the booting process and physically protects the enclave using the memory control registers. The program loaded in the enclave has the same CPU privilege as the OS kernel so that it can execute in parallel with the operating system. Hence, the attacker cannot access the enclave even if it can hack the OS or take control of the user application. Therefore, the program loaded into the enclave can execute in an isolated environment hence is invulnerable to software attack and DoS attack.

\subsubsection{True random number generation}
Random number generator (RNG) is frequently used by the cryptographic functions of the blockchain, particularly in private key generation, asymmetric encryption, and digital signature. However, the RNG provided by the OS is based on the pseudo-random number generation function that takes the system clock or software states as the random source, which is vulnerable to the software attack and side-channel attack. The attacker can control the output of the RNG by hacking the OS and controlling the value of these random sources. To address this issue, we implement a hardware-based true random number generator (TRNG) in the security chip. The TRNG utilizes the state of the hardware circuit (clock drift and thermal noise) as the random source that cannot be controlled by anyone. Furthermore, the execution of the TRNG is implemented directly in the circuit that the attacker cannot intervene.

\subsubsection{Hardware cryptographic module}
The cryptographic functions, especially the asymmetric cryptographic functions, lay the security foundation of the blockchain platform \cite{dai2019blockchain}. However, the cryptographic functions are also vulnerable to the software and side-channel attacks that can glean the used memory to derive the private key. To counter the attack vectors, we implement the common cryptographic functions (e.g., elliptic curve-based encryption and digital signature) as a hardware module in our security chip. The attackers can neither pry nor intervene the operation of the hardware cryptographic module; therefore, this module guarantees the security of the cryptographic functions at the hardware level.

\subsection{Trusted Software Framework}
Based on the trusted computing hardware discussed in the previous section, we further build a trusted software framework of the IIoT blockchain. 

\subsubsection{Root of trust}
The root of trust (RoT) refers to the most security-critical information that should be inherently trusted by the IIoT blockchain. Specifically, the RoT contains the endorsement key of the security chip and the private key of the blockchain nodes. The endorsement key provides a unique identity of the IIoT device and is used to generate the digital signature of the device. We harden the endorsement key into the security chip during the manufacturing process. The private key is generated during the first-time booting procedure and written into the non-volatile memory of the security chip. The private key is used to generate the address of the IIoT blockchain node and to sign digital signatures. The security chip guarantees that both the endorsement key and the private key are confidential and can only be accessed by the cryptographic module of the security chip. Therefore, the RoT module can be inherently trusted and safeguards the IIoT blockchain platform.

\subsubsection{Trusted blockchain}
To protect the blockchain from malicious attackers, we load the binary code of the blockchain software into the enclave created by the TEE module of the CPU as shown in Fig.~\ref{f3}. Because the enclave is an isolated execution environment, the blockchain is protected from both software attack and DoS attack. Comparing with running the blockchain software in the user mode, this approach has three security advantages. First, the proposed secure blockchain does not store any private key information in the memory; instead, it can obtain its trusted identity from the RoT module. Second, the secure blockchain node can sign transactions with a secure digital signature using the hardware cryptographic module, thus avoiding the software attack vectors. Third, the secure blockchain node can obtain true random numbers from the TRNG module and thus counters the random number generator attacks \cite{pinto2017iioteed}. ed

\begin{figure*}[!t]
    \centering
    \includegraphics[width=12cm]{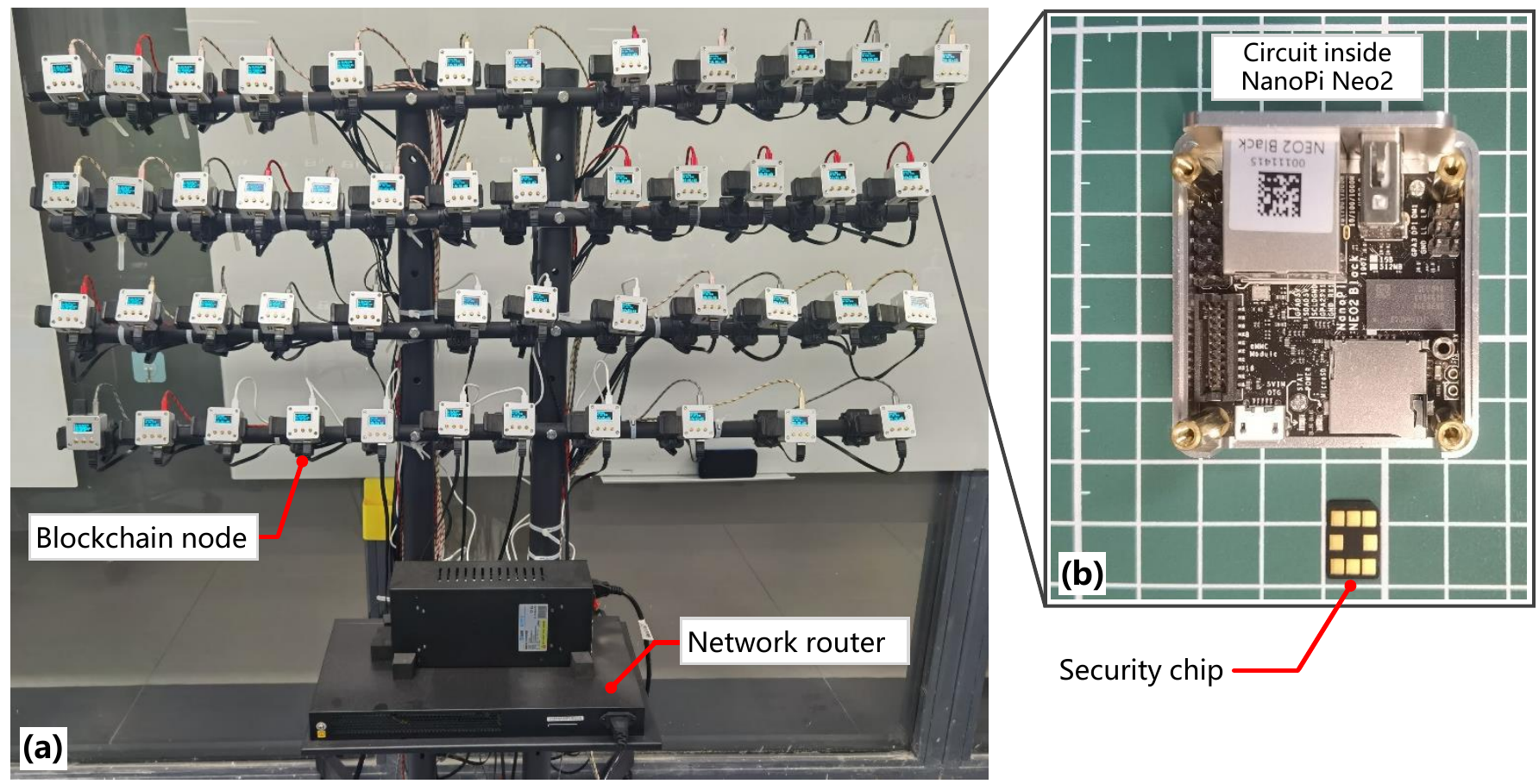}
    \caption{(a) The proof-of-concept IIoT blockchain network that consists of 48 NanoPi~Neo2 devices; (b) The circuit board of the NanoPi~Neo2 and the security chip.}
    \label{f4}
\end{figure*}

\subsubsection{Remote attestation}
In many IIoT applications, the binary code and the data of the application program should be intact in order to produce trusted results. For example, in the scenario of computation offloading in edge computing \cite{xu2019become}, a computing task is distributed to multiple IIoT devices. The distributed computing tasks require the IIoT node to perform a specific algorithm on certain data. However, malicious attackers may tamper with the binary code of the algorithm in the memory, since the user application resides in the unprotected memory. This attack leads to incorrect or even manipulated computing results and hence sabotages the distributed computing tasks. To address this issue, we implement a trusted remote attestation function in the TEE. The attestation function can attest to the binary code of the user application in the memory by calculating the hash of the code and generate an attestation report with the device's digital signature \cite{ambrosin2020collective}.

Specifically, the blockchain-based remote attestation mechanism proceeds in the following steps. First, the attestation function is loaded into the enclave during the secure booting procedure. Second, when publishing the distributed computing tasks, the attestation report of the binary code used in the task is also published on the blockchain. Third, when the IIoT device is executing the distributed computing task, the attestation function attests the computing task's binary code residues in the user space (in Fig.~\ref{f3}) and generates the attestation report; by comparing the local attestation report with that stored on the blockchain, the IIoT device knows whether the algorithm code is tampered or intact. In this way, the remote attestation function guarantees the integrity of the application software and data on the IIoT blockchain platform.

\rv{\subsection{Case Study}
We present a blockchain-based decentralized federated learning (FL) system for smart manufacturing that employs the proposed secure blockchain platform. In this FL system, multiple IIoT devices in different smart factories cooperatively train a global machine learning model with their local training data, which are privacy-sensitive. Instead of using a centralized coordinator to aggregate the local models, the decentralized FL system implements the model aggregation algorithm in a smart contract on the secure blockchain. The secure blockchain platform guarantees the correctness and trustness of the smart contract, thus removing the need for a centralized coordinator that may cause single-point failure. The secure blockchain platform running on IIoT devices can counter various security threats in the industrial environment. Furthermore, we use remote attestation to guarantee the integrity of the FL program run on the IoT devices. As a result, a secure and trusted decentralized FL system for IIoT applications is implemented by leveraging the proposed secure blockchain platform.}

\section{Implementation and Evaluation}\label{s:implementation}

This section presents the implementation and the performance evaluation of the proposed secure blockchain based on trusted computing hardware. \rv{We focus on two evaluation metrics of the secure blockchain platform, namely the security capability and the performance.} We first prototype the blockchain on IIoT devices to validate the design of the secure blockchain. Then, we build a small-scale IIoT network and evaluate the performance of the secure blockchain under different configurations.

\subsection{System Implementation}
As shown in Fig.~\ref{f4}, the NanoPi~Neo2 is a low-power embedded device with an ARM-A53 (quad-core $1.5$GHz) CPU and $1$GB memory, which are typical hardware configurations of IIoT devices. We package the NanoPi~Neo2 board with a metal housing and an OLED screen to display its status information. The bare circuit board of NanoPi~Neo2 and the customized security chip are shown in Fig.~\ref{f4}b. The security chip is packaged into a SIM card which can be inserted into the embedded device. The testing network consists of $48$ NanoPi~Neo2 nodes, which are connected to a network router via the Ethernet. 

We implement the design of the secure blockchain on the NanoPi~Neo2 device based on the source code of Ethereum release~1.7 \cite{geth}. Specifically, we modify the membership selection algorithm from PoW to PoS and adopt the PBFT consensus algorithm from the Hotstuff project \cite{yin2019hotstuff}. We offload the security-related functions of the blockchain to the security chip. In our test, the secure blockchain software consumes about $180$MB memory and $100$MB memory for a validator node and a non-consensus node, respectively. All the blockchain nodes run the Ubuntu Core 16.04 as the operating system. We utilize the TrustZone \cite{ngabonziza2016trustzone} in the ARM CPU as the TEE module and create an enclave in the memory to host the blockchain software and the remote attestation function.

\begin{figure*}[!t]
    \centering
    \includegraphics[width=11cm]{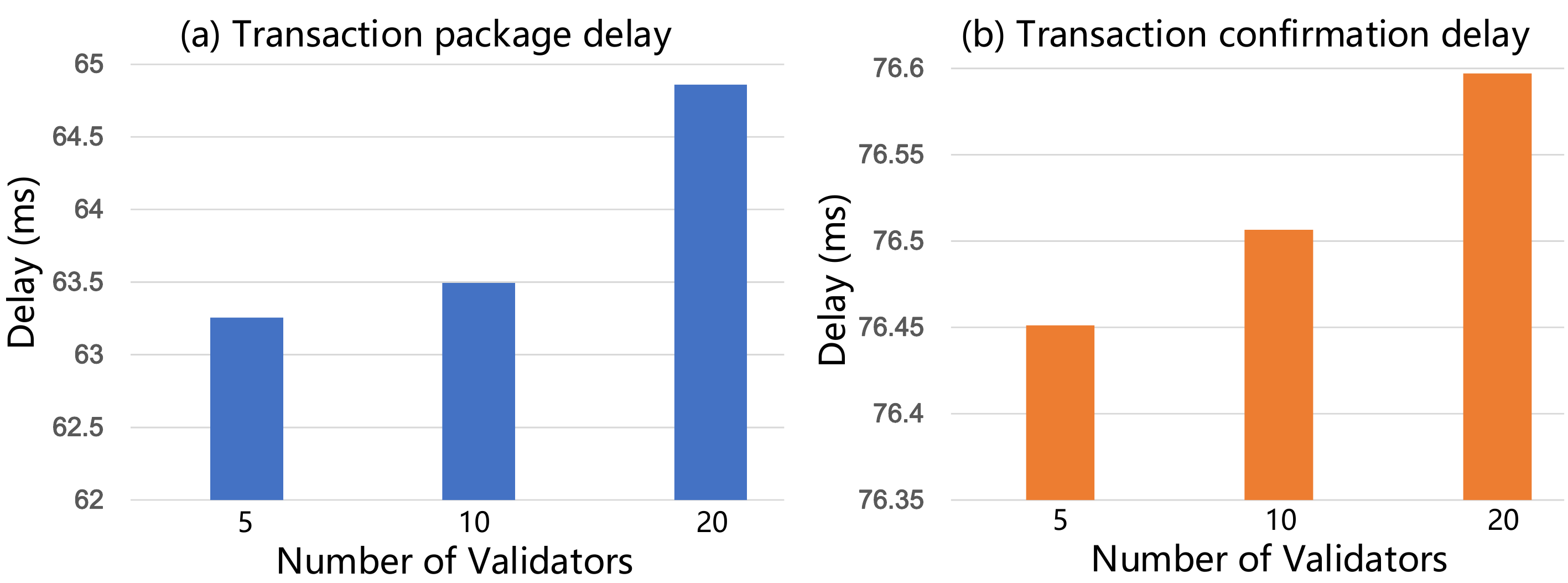}
    \caption{(a) The average transaction processing delay. (b) The average transaction confirmation delay. The number of PBFT validators varies from 5 to 20.}
    \label{f5}
\end{figure*}

\begin{figure*}[!t]
    \centering
    \includegraphics[width=11cm]{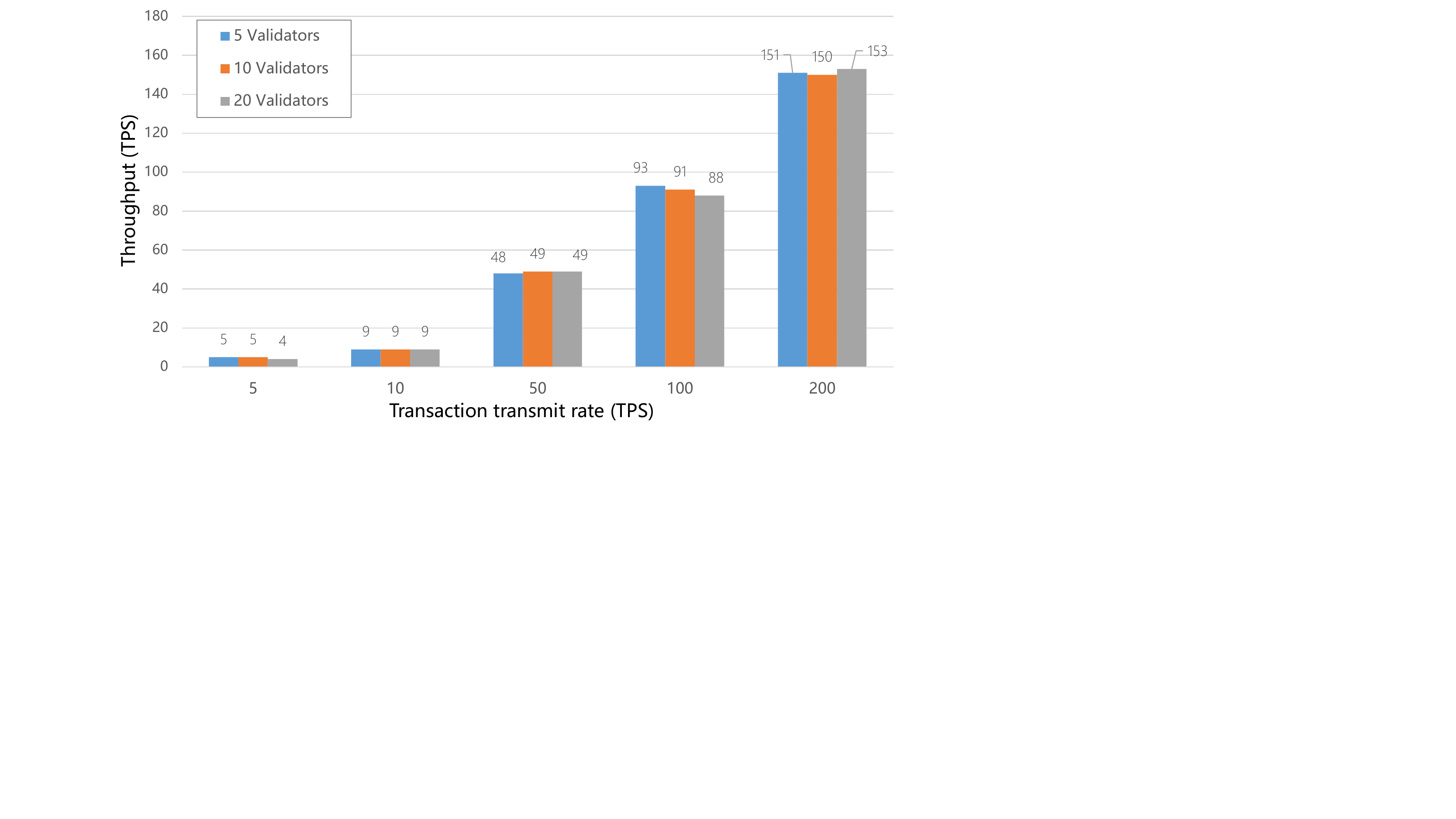}
    \caption{The throughput of the secure IIoT blockchain platform measured in transaction per second (TPS).}
    \label{f6}
\end{figure*}

\subsection{Security Analysis}
\rv{We analyze and compare the secure capability of different IIoT platforms, including the proposed secure blockchain, cloud computing, and normal blockchain such as Ethereum. Unlike Ethereum and cloud computing, the secure blockchain program is running in the enclave protected by the TEE module; therefore, the secure blockchain is invulnerable to the software attack and DoS attack. Normal blockchains are vulnerable to hardware and side-channel attacks because the IIoT nodes are usually exposed in the environment. By contrast, all the cryptographic operations of the secure blockchain platform (e.g., random number generation, key management, asymmetric encryption) are offloaded to the security chip, thus resists the hardware attack and side-channel attack. Furthermore, we employ remote attestation to counter the software attack and guarantee the software integrity of the off-chain applications in the user space, such as data acquisition and fog computing. As a result, the secure blockchain platform can effectively cope with the security threats in IIoT scenarios.}

\subsection{Performance Evaluation}

We evaluate the performance of the secure blockchain platform in the testing network of $48$ IIoT nodes. In the experiment, we consider three cases with $5$, $10$, and $20$ PBFT validators (consensus nodes), respectively. In each case, we let the non-consensus nodes transmit transactions at a fixed transmit rate ranging from $5$ to $200$ transactions per second (TPS). To measure the efficiency of the secure blockchain platform, we measure two kinds of delays: 1) the \emph{transaction processing delay} measures the time that elapses from the moment when the node transmits a transaction to the moment when this transaction is packaged into a block by the leader validator; 2) the \emph{transaction confirmation delay} measures the time elapses from the moment when the node transmits a transaction to the moment when the block containing this transaction is decided by the consensus algorithm. The measured delays are illustrated in Fig.~\ref{f5}.

In Fig.~\ref{f5}a, we observe that the transaction package delay slightly increases when the number of validators increases, but the delay value are quite close and below $65$ms. This result indicates that the transaction processing procedure of the validators is efficient. Fig.~\ref{f5}b shows that the transaction confirmation delays with different numbers of validators are also close and below $76.6$ms. Comparing Fig.~\ref{f5}a and Fig.~\ref{f5}b, we can estimate that the time consumed by the consensus algorithm is below $10$ms, which validates the feasibility and efficiency of the secure blockchain.

The throughput performance of the secure blockchain measured in TPS is shown in Fig.~\ref{f6}. In the experiment, we gradually increase the transaction transmit rate and measure the average number of confirmed transactions by the consensus algorithm. When the transmit rate is low (e.g., $5$, $10$, and $50$TPS) the throughput is equal to the transmit rate, indicating that all the transactions are confirmed in time; however, when the transmit rate is high (e.g., $100$ and $200$TPS), the processing capability of the consensus algorithm is saturated so that some transactions are delayed. The peak throughput of the proposed secure blockchain platform in the testing network is about $150$TPS, which is much higher than that of Bitcoin ($7$TPS) and Ethereum ($28$TPS). The experimental results show that the proposed secure blockchain platform has the potential to support various IIoT applications in practice.

\section{Future Research Directions}
\label{s:future}
During the investigation of this work, we found several open issues worth further exploration in the future.

\textbf{New TEE architecture for IIoT blockchain}: According to our study of this paper, we find that 
employing TEE can simplify the design and enhance the security level of the blockchain system. The existing TEEs provided by the major CPU vendors, however, have only limited and inflexible functions that cannot meet the requirements of secure blockchain systems. For example, the ARM TrustZone only supports a single TEE enclave, and the memory size of the Intel SGX's TEE enclave is limited to $128$MB. Existing TEE designs also lack the support of cryptographic functions and TRNG, so we have to introduce a customized security chip in this work. Recently, the emerging of open-source CPUs (e.g., RISC-V) makes it possible for users to design customizable TEE frameworks \cite{lee2020keystone}, providing a new research direction for TEE-based secure blockchain. TEE customized for IIoT blockchain can offload security-critical functions, facilitate consensus algorithms, and safeguard blockchain applications. Therefore, designing new TEE architecture for IIoT blockchain is an interesting future research direction.

\textbf{Privacy protection in IIoT scenarios}: One drawback of this work is lacking consideration of privacy protection. The blockchain is a distributed ledger on which all the transactions and data are transparent, hence incurs the risk of privacy leakage. We focus on identity privacy and data privacy in IIoT scenarios. First, although the blockchain adopts the pseudo-anonymity mechanism to hide the user's identity, attackers can still infer the user's real identity by analyzing the linkage among the transactions related to the user's address \cite{khan2018iot}. Second, the user's data in smart contracts are all public and transparent on the blockchain, which may contain the user's private information. These privacy issues violate the privacy-protection laws (e.g., the General Data Protection Regulation) and also limit the application of the IIoT blockchain. Recent researches on zero-knowledge proof (ZKP) provides an effective method to issue transactions without leaking any identity information. Furthermore, homomorphic encryption, which allows users to perform computations on encrypted data without decrypting it, can be used to achieve privacy-preserving computation in smart contracts. Integrating ZKP and homomorphic encryption into the IIoT blockchain to protect users' privacy in IIoT scenarios remains a promising research area.

\textbf{High-performance blockchain for IIoT devices}: Most blockchain systems are originally designed to run on computers that have plenty of hardware resources (CPU, storage, memory) and network bandwidth. The IIoT device, however, usually has limited hardware resources and bandwidth due to its size and installation environment; hence the performance of existing blockchains degrades severely on IIoT devices. We identify three potential directions to build a high-performance blockchain for IIoT devices. First, new consensus algorithms with low complexity \rv{are worth investigating}. The mainstream consensus algorithms are infeasible for IIoT devices because they are either computation-intensive (e.g., PoW) or communication-intensive (e.g., PBFT) that requires high computing power or network bandwidth. Second, efficient message transmission protocol should be studied to lower the requirement of network bandwidth. The Gossip protocol, which is adopted by mainstream blockchains, uses the epidemic algorithm to disseminate messages in the network. This method incurs unnecessary message redundancy and wastes network bandwidth, so it is unsuitable for IIoT networks. \rv{Third, scaling strategies to improve the throughput of the blockchain on IoT devices are of great importance. We will explore promising scaling techniques such as sharding, side chain, and hardware-assisted acceleration. Further study along these directions is ongoing in our lab to implement a high-performance blockchain for IIoT devices.}

\section{Conclusion}\label{s:conclusion}
This paper presented a secure blockchain platform based on trusted computing hardware for industrial IoT (IIoT). Specifically, we employed the trusted execution environment (TEE) and a customized security chip in the blockchain design to counter various security threats in IIoT scenarios. We first designed the architecture of the secure blockchain from the aspects of membership selection, consensus algorithm, and blockchain structure. Further, we elaborated on the trusted computing hardware and trusted software framework that safeguards the blockchain platform from malicious attack vectors. To validate the proposed design, we implemented and evaluated the secure blockchain on a testing network of $48$ NanoPi nodes that resemble a practical IIoT system. The experimental results showed that the secure blockchain platform can achieve low transaction confirmation delay and high throughput, validating its feasibility in real IIoT networks. Finally, we discussed the open issues that lead to the future research direction. The investigation and primary results in this work suggest that secure blockchain based on trusted computing hardware is a promising solution to facilitate the application of blockchain in practical IIoT scenarios.

\bibliographystyle{IEEEtran}
\bibliography{ref.bib}

\end{document}